\documentclass[a4,printer]{aa}  
\usepackage{graphicx}
\usepackage{txfonts}
\begin{document}
   \title{Not an open cluster after all: the NGC 6863 asterism in Aquila
          \thanks{Based on observations carried out at ESO La Silla, 
                  under program 281.D-5054}
         }
   \titlerunning{The NGC 6863 asterism}

   \author{C. Moni Bidin$^{1}$, R. de la Fuente Marcos$^{2}$, 
           C. de la Fuente Marcos$^{2}$, and G. Carraro$^{3,4}$} 
   \authorrunning{Moni Bidin et al.}

   \offprints{Giovanni Carraro: gcarraro@eso.org}

   \institute{$^1$Departamento de Astronomia, Universidad de Concepcion, 
                  Casilla 160-C, Concepcion, Chile\\
              $^2$Suffolk University Madrid Campus, C/ Vi\~na 3, 
                  E-28003 Madrid, Spain\\
              $^3$European Southern Observatory, Alonso de Cordova 3107, 
                  Vitacura, Santiago, Chile\\
              $^4$Dipartimento di Astronomia, Universit\'a di Padova, 
                  vicolo Osservatorio 3, I-35122, Padova, Italy\\
                  }

   \date{Received June 19, 2009; accepted XXXXXXX XX, XXXX}
 
   \abstract
       {Shortly after birth, open clusters start dissolving; gradually 
        losing stars into the surrounding star field. The time scale for 
        complete disintegration depends both on their initial membership 
        and location within the Galaxy. Open clusters undergoing the 
        terminal phase of cluster disruption or open cluster remnants 
        (OCRs) are notoriously difficult to identify. From an observational
        point, a combination of low number statistics and minimal contrast 
        against the general stellar field conspire to turn them into very 
        challenging objects. To make the situation even worst, random 
        samples of field stars often display features that may induce to 
        classify them erroneously as extremely evolved open clusters.} 
       {In this paper, we provide a detailed study of the stellar content 
        and kinematics of NGC~6863, a compact group of a few stars located
        in Aquila and described by the Palomar Observatory Sky Survey as a 
        non existent cluster. Nonetheless, this object has been recently 
        classified as OCR. The aim of the present work is to either confirm 
        or disprove its OCR status by a detailed star-by-star analysis.} 
       {The analysis is performed using wide-field photometry in the UBVI 
        pass-band, proper motions from the UCAC3 catalogue, and high 
        resolution spectroscopy as well as results from extensive $N$-body 
        calculations.}
       {The spectra of the four brightest stars in this field clearly 
        indicate that they are part of different populations. Their radial 
        velocities are statistically very different and their spectroscopic 
        parallaxes are inconsistent with them being part of a single, bound 
        stellar system. Out of the four stars, only two of them have similar 
        metallicity. The color magnitude diagram for the field of NGC~6863 
        does not show any clear signature typical of actual open clusters.
        Consistently, spatial scan statistics confirms the absence of any
        statistically significant, kinematically supported over-density at 
        the purported location of NGC~6863.} 
       {Our results show that the four brightest stars commonly associated
        to NGC~6863 form an asterism, a group of non-physically associated 
        stars projected together, leading to the conclusion that NGC 
        6863 is not a real open cluster.}  


   \maketitle
   \section{Introduction}
      Most if not all the field stars that we observe on any given galaxy 
      are thought to have been formed in star clusters or associations 
      (see e.g. Lada \& Lada 2003). In the Milky Way disk and as soon as 
      they come into existence these objects become dissolving clusters, 
      gradually losing stars into the general field population. Open 
      clusters are known to disintegrate on a time scale which depends both 
      on their membership and galactocentric distance. The eventual residue 
      of the evolution of an open cluster is often called open cluster 
      remnant (hereafter OCR). Early $N$-body simulations gave us a first 
      glimpse on how the final stage in the life of an open cluster may look 
      like. For simulated systems with some 25 to 250 stars, von Hoerner 
      (1960, 1963), Aarseth (1968) and van Albada (1968) suggested that the 
      final outcome in the evolution of an open star cluster is one or more 
      tightly bound binaries (or even a hierarchical triple system). Van 
      Albada pointed out several observational candidates ($\sigma$ Ori, 
      ADS 12696, $\rho$ Oph, 1 Cas, 8 Lac and 67 Oph) as being OCRs and 
      Wielen (1975) indicated yet another one, the Ursa Major cluster 
      (Collinder 285). 

      Regrettably, these early numerical results failed to spur interest 
      among the observational community and trigger any systematic study 
      aimed at identifying open clusters undergoing the terminal phase of 
      cluster disruption: the OCR stage. We had to wait until the late 80s 
      to find the first attempt. Using objective-prism plates, Lod\'en 
      (1987, 1988, 1993, and references therein) investigated the frequency 
      of clusters in the Milky Way under the hypothesis that their stars 
      should have similar luminosities and spectral types. He found that 
      about 30\% of the objects in his sample could be catalogued as a 
      possible type of cluster remnant. Unfortunately, without proper 
      kinematic information, this approach produces non conclusive results. 
      There are known regions in the sky with many stars of similar spectral 
      type but in which it is difficult to find two stars with the same 
      velocity. These are some type of asterism: their stars are all at 
      different distances and move in different directions but perspective 
      lines them up in the sky to become optical groupings. The situation is 
      analogous to optical doubles and binary stars. New simulations (de la 
      Fuente Marcos 1997, 1998) confirmed early numerical results 
      characterizing OCRs as sparsely populated stellar aggregates, rich in 
      binaries and higher multiplicity systems and poor in low-mass members. 
      Another main conclusion of these simulations was the intrinsically 
      specific stellar content of remnants coming from clusters with very 
      different initial populations. Open clusters originally hosting 
      relatively small numbers of stars (perhaps not more than 10$^3$) still 
      arrive to the terminal phase of cluster evolution with a significant 
      population of early-type, luminous stars. Enhanced preferential retention
      of massive stars and binaries is mainly the result of dynamically 
      induced mass segregation. In contrast, originally rich clusters (e.g. 
      $N > 10^4$) can survive for several Gyr and when they arrive to the 
      twilight of their dynamical lives, the brightest stars still in the main 
      sequence are late F or early G spectral types. In other words,
      initially richer clusters remain bound for longer time and therefore
      the main sequence tunoff point is fainter by the time that they
      dissolve. This fact makes OCRs from these clusters much more 
      difficult to identify than those from smaller clusters.
       
      Unconnected to Lod\'en's work, a new list of OCR candidates was produced 
      by Bica et al. (2001); their members being selected based on stellar 
      density contrast considerations. Unfortunately, subsequent analyses 
      including positional, kinematic, photometric and spectroscopic 
      information failed to confirm even the cluster status of some of the 
      suggested candidates (Villanova et al. 2004a; Carraro et al. 2005). 
      Again, negative results pointed out an already well known fact, that 
      these objects are notoriously difficult to identify. This relatively 
      large fraction of false positives strongly hints that random samples of
      field stars often display features that may induce to classify them 
      erroneously as extremely evolved open clusters. In spite of their 
      intrinsically challenging nature, the observational search for OCR 
      candidates has become a subject of considerable interest as it is one 
      of the keys to understand the origin of the field star populations of 
      the Galactic disk (see Carraro 2006). 

      In this research, we provide a detailed study of the stellar content 
      and kinematics of NGC~6863, a compact group of a few stars in Aquila 
      described by the Palomar Observatory Sky Survey as a non existent 
      cluster. However, this object has been recently classified as OCR 
      (Bica et al. 2001; Pavani \& Bica 2007). The aim of the present work 
      is to either confirm or disprove this result. In this paper, we present 
      new CCD UBVI photometry and single epoch spectroscopy, which we combine 
      with proper motions from the UCAC3 (Zacharias et al. 2009) catalogue in
      an attempt to unravel the true nature and dynamical status of this 
      object. This paper is organized as follows: in Sect. 2 we provide a 
      historical overview of the data previously available for this object. 
      Observations, data reduction, and overall results are presented in 
      Sect. 3. In Sect. 4 we focus on proper motions. The statistical 
      significance of the results is discussed in Sect. 5. In Sect. 6 we draw 
      our conclusions.
 
%
%
    \begin{figure}
     \centering
      \includegraphics[width=8.5cm]{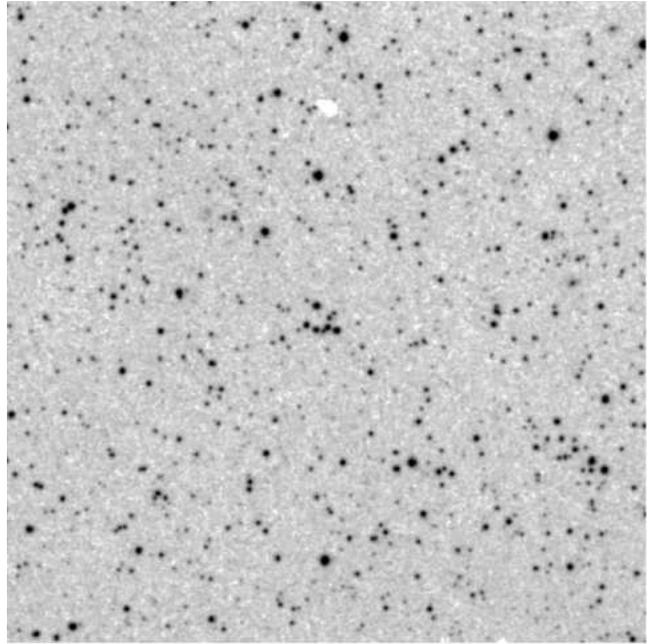}
      \caption{Palomar Observatory Sky Survey (POSSI) image of the area 
               covered by the present study. North is up, East to the 
               left, and the image is 12.9 arcmin on a side. This 12 min
               digitized plate was acquired by the Palomar 48-in Schmidt 
               camera using the red (O) filter on 1951.50678. The
               frame displayed has been processed using square root 
               scaling.}
      \label{ngc6863}
    \end{figure}
%
%

   \section{NGC 6863 in perspective}
      This asterism or group of a few stars in Aquila was first identified in
      1827 by Herschel (1833) as h 2065 using a 18.7 inches telescope. The same
      group of eight Galactic stars was included as object GC 4542 by Herschel 
      (1864) in his "Catalogue of Nebulae and Clusters of Stars". The "New 
      General Catalogue of Nebulae and Clusters of Stars" (Dreyer 1888) first 
      used the name NGC 6863, describing the object as a very much compressed, 
      small cluster including stars from the 19th magnitude downwards. No 
      further references to this object are found until the publication of 
      "The Revised New General Catalogue of Nonstellar Astronomical Objects" 
      (Sulentic \& Tifft 1973). This catalog is a modern, revised, and 
      expanded version of the original NGC. Besides incorporating the many 
      corrections to the NGC found over the years, each object was verified on 
      Palomar Observatory Sky Survey (POSS) prints and on plates for southern 
      objects specifically taken for the purpose. In this catalog NGC 6863 is 
      described as non existent. The POSSI red (O) plate is displayed in Fig. 
      \ref{ngc6863}. The object is widely regarded as an asterism among the 
      amateur astronomical community (see, e.g., Allison 2006) but it is listed
      as a {\it bona fide} cluster in both the {\it Open Cluster Database} 
      \footnote{http://www.univie.ac.at/webda/} (WEBDA, Mermilliod \& Paunzen 
      2003) and the {\it New Catalogue of Optically Visible Open Clusters and 
      Candidates} \footnote{http://www.astro.iag.usp.br/$\sim$wilton/} (NCOVOCC,
      Dias et al. 2002). These databases are widely used in professional open 
      cluster studies. The latest update of WEBDA (November 2009, Paunzen \& 
      Mermilliod 2009) includes coordinates, 
      $\alpha=20^{\rm h}~05^{\rm m}~07^{\rm s}$, $\delta=-03^{\circ} 
      33^{\prime}18^{\prime\prime}$, $l=38^{\circ}.278$, $b=-17^{\circ}.996$, 
      J2000, and diameter, 2 arcmin, for NGC 6863 but no other data are 
      provided. The February 2009 version (v2.10, Dias 2009) of NCOVOCC 
      includes NGC 6863 as a well established open cluster with coordinates, 
      $\alpha=20^{\rm h}~05^{\rm m}~07^{\rm s}$, $\delta=-03^{\circ} 
      33^{\prime}20^{\prime\prime}$, $l=38^{\circ}.2778$, $b=-17^{\circ}.9962$, 
      J2000, and diameter, 3 arcmin, located 1200 pc from the Sun and with an 
      age of 3.47 Gyr. SIMBAD \footnote{http://simbad.u-strasbg.fr/simbad/} 
      also classifies NGC 6863 as cluster of stars and refers to Bica et al. 
      (2001). Conversely, the NGC/IC Project 
      \footnote{http://www.ngcicproject.org} 
      (Erdmann 2009) indicates that NGC 6863 is an asterism made of 8 stars.

      An extensive search across published literature on this object reveals 
      only three entries, all of them within the context of the Open Cluster 
      Remnant paradigm. Bica et al. (2001) included NGC 6863 in their list of 
      dissolving star cluster candidates with Galactic coordinates 
      $l=38^{\circ}.27$, $b=-17^{\circ}.99$, $E(B-V)$ = 0.28, and age 3-4 Gyr. 
      Pavani et al. (2007) presented preliminary spectroscopy for NGC 6863. 
      Finally and using Two Micron All Sky Survey (2MASS, Skrutskie et al. 1997)
      and UCAC2 (Zacharias et al. 2004) data, Pavani \& Bica (2007) studied the
      structure, proper motions, and CMD distribution of this object to conclude
      that NGC 6863 is an open cluster remnant of old age, 3.5$\pm$0.5 Gyr, its
      centre and characteristic size as in Dias (2009), see above. An obvious 
      over-density of bright stars can be seen in the central region of Fig. 
      \ref{ngc6863}; however and following (e.g.) Odenkirchen \& Soubiran 
      (2002), an overdensity of stars does not necessarily imply the existence 
      of a physical ensemble.
        
   \section{Observations and reduction}
      Observations were performed in two different runs. The photometric one was carried 
      out at Asiago (Italy) on December 2001, whereas the spectroscopic run was completed 
      at ESO La Silla on September 2008. Details of both runs are given in the following 
      sections.

      \subsection{Photometry}
         Observations were carried out with the AFOSC camera at the 1.82~m Copernico 
         telescope of Cima Ekar (Asiago, Italy), in the photometric nights of December 17 
         and 18, 2001. AFOSC samples a $8^\prime.14\times8^\prime.14$ field in a $1K\times1K$ 
         thinned CCD. The typical seeing was between 1.8 and 2.3 arcsec. The details of the 
         observations and data reductions are provided in Villanova et al. (2004a), where we 
         reported on NGC~5385, Collinder~21 and NGC~2664, which were observed in the same run. 
         Basic information on the NGC~6863 photometric run is given in Table \ref{photo}.

\begin{table}
 \caption{Journal of photometric observations in the field of NGC~6863} 
  \begin{tabular}{cccc} 
   \hline 
    \multicolumn{1}{c}{Field}    & 
    \multicolumn{1}{c}{Filter}    & 
    \multicolumn{1}{c}{Time integration}& 
    \multicolumn{1}{c}{Seeing}         \\ 
               &     &  (sec)        & ($\prime\prime$)\\ 
    \hline 
     NGC~6863  &     &               &      \\ 
               & $U$ &  15, 300      &  1.8 \\ 
               & $B$ &  2, 5, 90     &  1.9 \\ 
               & $V$ &  1, 3, 30, 30 &  1.7 \\ 
               & $I$ &  1, 5, 30     &  1.6 \\ 
    \hline 
  \end{tabular} 
  \label{photo}
\end{table} 

    \begin{figure}
     \centering
      \includegraphics[width=9cm]{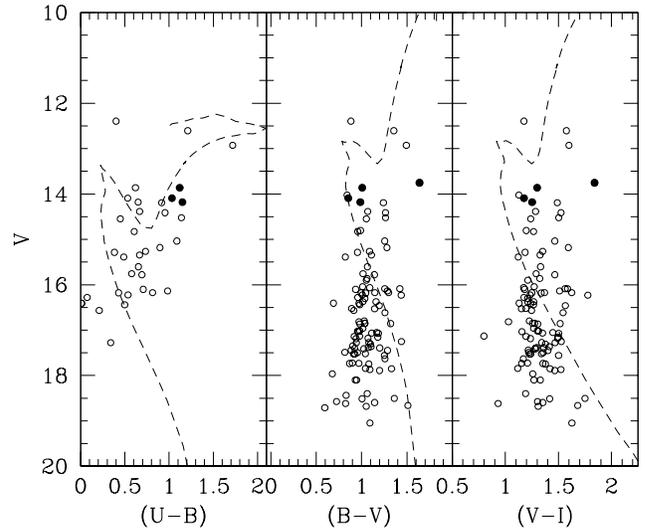}
      \caption{CMDs in various color combinations for all the stars having UBVI photometry
              in the field of NGC~6863. Filled simbols indicate the 4 stars which identify
              the over-density (see Fig. \ref{ngc6863}). Super-imposed are a solar metallicity
              isochrone from Girardi et al. (2000) adopting the set of parameters derived
              by Pavani et al. (2007) for this asterism.}
      \label{center}
      \label{ngc6863cmd}
    \end{figure}

         From our photometric data we construct color-magnitude diagrams (CMD) for several 
         color combinations.  They are shown in Fig. \ref{ngc6863cmd}. Here and with filled 
         symbols we indicate the stars defining the asterism at the centre of the NGC~6863 
         field. These four stars produce the appearance of an overdensity in DSS images.
         All of them have UBVI measurements with the exception of star \#4, which does not 
         have U. The CMDs for the detected stars clearly do not show any 
         distinctive feature that can lead us to think about a group of stars with common 
         properties, a physical ensemble as in the case of, e.g., NGC~1901 discussed in 
         Carraro et al. (2007). To guide the eye, and better clarify our conclusion,
         we over-impose in the same Figure a solar metallicity isochrone taken from
         the Padova database (Girardi et al. 2000), and adopting the set of parameters
         from Pavani et al. (2007): 3.5 Gyrs for the age, E(B-V) = 0.28, and
         (m-M)$_o$ $\sim$ 10.0. 

         Our photometric material simply suggests that what we see in 
         the direction of NGC~6863 is nothing but random Galactic field.

      \subsection{Spectroscopy}
         The spectra of the four brightest stars in NGC\,6863 were collected at La Silla 
         Observatory on 2008, September 18, with the HARPS fibre-fed spectrograph at the 
         Cassegrain focus of the 3.6m telescope. Exposure times varied between 1800s and 3600s 
         depending on target brightness, and the resulting signal-to-noise ratio for each 
         spectrum is given in column~2 of Table~\ref{spectradata}. The light contribution from 
         sky background was not negligible for our faint targets, and we allocated the second 
         fiber to the sky instead of the simultaneous wavelength-calibration lamp, thus relying
         for calibration only on lamp frames acquired during daytime operations. However, the 
         very high spectral resolution and instrumental stability guaranteed an accuracy still 
         much higher than what was required for our aims.

\begin{figure}
  \begin{center}
    \resizebox{\hsize}{!}{\includegraphics{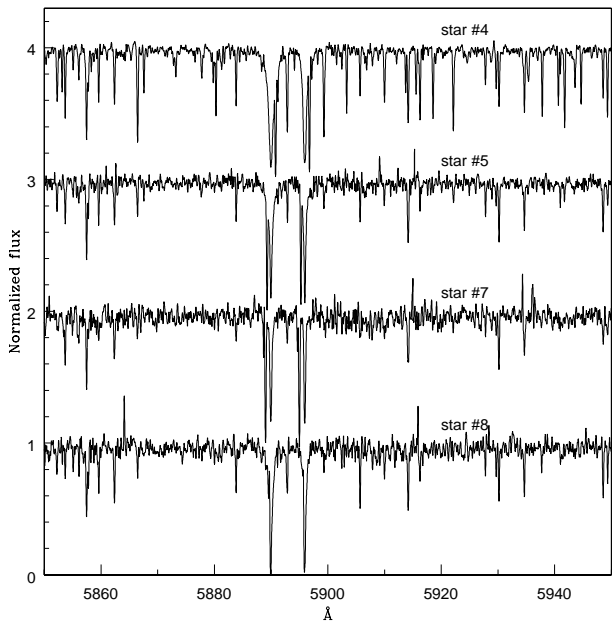}}
      \caption{A portion of the collected spectra, centred on the strong NaI doublet. 
               An integer constant have been added to shift spectra vertically.}
      \label{fig_spectra}
  \end{center}
\end{figure}

         We first attempted a manual data reduction by means of standard IRAF\footnote{IRAF is 
         distributed by the National Optical Astronomy Observatories, which are operated by the 
         Association of Universities for Research in Astronomy, Inc., under cooperative agreement 
         with the National Science Foundation.} routines, comprising all steps from bias and 
         flat-field corrections to spectra extraction and wavelength calibration. The inspection 
         of the results revealed that there was no improvement, in terms of S/N and spectral 
         quality, with respect to the output of the online pipeline at the telescope. The wavelength
         calibration was also equivalent within calibration errors, and no systematic effect was 
         detected. Hence, we adopted the pipeline-reduced, extracted and calibrated spectra. The 
         extracted sky spectrum from the second fiber was subtracted to each target. The order \#46 
         of the second fiber falls between the two CCDs, and the corresponding science order was 
         also suppressed, generating a gap between 5258 and 5336 ~\AA. Spectra were convolved with 
         a Gaussian filter to decrease the resolution to about R=38\,000, then re-binned in blocks of 
         4 pixels and re-sampled to a constant wavelength step of 0.01 ~\AA. We thus increased the 
         S/N by a factor of two, while still keeping the resolution required for precise radial
         velocity and abundances measurements. Resulting spectra were divided order-by-order by the 
         blaze function obtained by the pipeline from flat field frames, and any residual curvature 
         of orders was eliminated fitting the continuum with a second-order polynomial. The orders 
         thus normalized were finally merged to produce the final spectra. A section of the final 
         spectra across the strong NaI doublet (5890-5893 \AA) is presented in Fig.~\ref{fig_spectra}.
         The whole useful spectral range goes from 4000 to 6800\AA. 

         \subsubsection{Radial velocities}
            Radial velocity (RV) measurements were performed with a cross-correlation (CC) technique 
            (Tonry \& Davis 1979) using the {\it fxcor} IRAF task. As template we used a synthetic spectrum 
            of solar metallicity drawn from the library of Coelho et al. (2005), with temperature and 
            surface gravity intermediate between the values expected for program stars (4750~K and 2.0 
            dex, respectively). We verified through repeated measurements that different assumptions on 
            template parameters in the range T$_\mathrm{eff}$=4000-5500~K and $\log{g}$=1.0-3.0 caused
            negligible changes in the resulting velocities ($\leq$0.2 km s$^{-1}$). The whole useful 
            spectral range was used in the CC, excluding the gap at 5300~K and the 
            head of the telluric band at the red end of the range. In order to fix the zero-point of 
            the measurements, this band was cross-correlated with a twilight solar spectrum acquired 
            during a different run. Within 0.1 km s$^{-1}$ the CC returned the solar RV at the dates of
            observations\footnote{http://eclipse.gsfc.nasa.gov/TYPE/TYPE.html}, indicating that no 
            correction for systematic effects was required. Finally, observed RVs were corrected to 
            heliocentric velocities. Errors have been estimated as the quadratic sum of the most relevant 
            sources, i.e. CC error (0.2-0.4 km s$^{-1}$), the uncertainty in zero-point definition (0.1 
            km s$^{-1}$), and choice of the synthetic template (0.2 km s$^{-1}$). Results are given in 
            column~7 of Table~\ref{spectradata}.

\begin{table*}
\caption{Stellar data derived with spectroscopic analysis}
\label{spectradata}
\begin{tabular}{ccccccccc}
\hline
ID & $\frac{S}{N}$ & T$_\mathrm{eff}$ & $\log{g}$ & [$\frac{\mathrm{Fe}}{\mathrm{H}}$] & $v_t$ & RV & $E(B-V$) & d \\
 & & K & dex & dex & km s$^{-1}$ & km s$^{-1}$ & mag & pc \\
\hline
4 & 25 & 4110$\pm$65 & 1.05$\pm$0.20 & $-$0.62$\pm$0.08 & 1.28$\pm$0.20 & -55.1$\pm$0.3 & 0.23$\pm$0.05 & 6800$\pm$1100 \\
5 & 15 & 5440$\pm$65 & 3.55$\pm$0.20 & $-$0.09$\pm$0.08 & 0.65$\pm$0.20 &  21.6$\pm$0.5 & 0.30$\pm$0.05 & 1270$\pm$280 \\
7 & 10 & 6100$\pm$75 & 4.45$\pm$0.20 & $-$0.06$\pm$0.09 & 1.32$\pm$0.15 &  35.3$\pm$0.4 & 0.34$\pm$0.04 &  560$\pm$50 \\
8 & 10 & 5320$\pm$65 & 3.85$\pm$0.10 & $-$0.41$\pm$0.10 & 1.24$\pm$0.15 & -10.6$\pm$0.5 & 0.26$\pm$0.05 &  740$\pm$90 \\
\hline
\end{tabular}
\end{table*}

         \subsubsection{Atmospheric parameters}
            Effective temperature (T$_\mathrm{eff}$), surface gravity ($\log{g}$), and metallicity of stars 
            \#4 and \#5 were measured with the ionization and excitation balance method. In brief, equivalent 
            widths (EWs) of a great number of FeI and FeII lines are measured, then T$_\mathrm{eff}$ and 
            microturbulent velocity ($v_t$) are determined imposing that FeI abundances ($\epsilon$(FeI))
            of each individual line show no trend with excitation potential (EP) and EW, respectively, while 
            $\log{g}$ is fixed by the requirement that FeI and FeII lines return the same iron abundance. We 
            adopted the general iron line list of Moni Bidin (2009), which comprises atomic parameters for 
            about 180 FeI and 40 FeII lines, with solar oscillator strengths and a line-by-line definition of 
            the correction term to the Uns\"old (1955) approximation of the collisional line broadening. However, 
            we rejected the measurement when the line fit was not satisfactory or the stellar continuum was not 
            easily defined, and only $\sim$70 iron lines were finally used for each star. EWs were measured with 
            a Gaussian fit with a dedicated IRAF script. Iron abundances were calculated with the LTE code 
            MOOG\footnote{freely distributed by C. Sneden, University of Texas, Austin} (Sneden 1973), with model 
            atmospheres obtained interpolating from the Kurucz (1992) grid. The results are given in 
            Table~\ref{spectradata}. We adopt the error estimates of Moni Bidin (2009), who compared the results 
            with literature values for thirteen stars with multiple previous measurements. However, we improved 
            the problematic fit of FeII lines, found at shorter wavelengths and affected by heavy line crowding, 
            by simultaneously fitting all features in a region centred on the line of interest. Measurements on 
            some test stars indicate that the scatter of $\log{g}$ is thus reduced, and we adopt here a corresponding
            lower error.

            EW measurement for stars \#7 and \#8 was possible for too few lines ($\leq$30) and with large 
            uncertainties, because of the low S/N. Consequently, the atmospheric parameter determination failed. For 
            example, for star \#7 there was no physically meaningful combination of parameters that permitted to zero 
            the trend of $\epsilon$(FeI) with EWs and EP simultaneously. We therefore elaborated a different strategy. 
            The wings of Balmer lines are a good indicator of temperature in the range T$_\mathrm{eff}$=5000-6500~K
            (Fuhrmann et al. 1994), and an explorative inspection of synthetic spectra revealed that they are quite 
            insensitive to $\log{g}$. We fitted the wings of H$_{\alpha}$ on target spectra with synthetic templates 
            at different temperatures in steps of 250~K, and results were fitted using a second-order polynomial to 
            determine the minimum of the $\chi^{2}$-T$_\mathrm{eff}$ relation. The estimate was repeated within a grid
            of values of metallicity and gravity ($\log{g}$ from 1 to 4, [Fe/H]=0 and $-$0.5) to check the extent of 
            systematics, but only small random differences were found. We will assume the mean of these repeated 
            measurements as the final result, and their rms as the associated error. We first tested the procedure on 
            star \#1, obtaining 5400$\pm$65~K. The excellent agreement with our estimate obtained with the EW method 
            indicates that the results are reliable. It was not possible to repeat the test on star \#4, because  
            Balmer lines rapidly vanish at decreasing temperature, and their reduced wings become insensitive to 
            temperature changes. In fact, for this cool star the $\chi^{2}$-T$_\mathrm{eff}$ was flat below about 4300~K.

\begin{figure}
  \begin{center}
    \resizebox{\hsize}{!}{\includegraphics{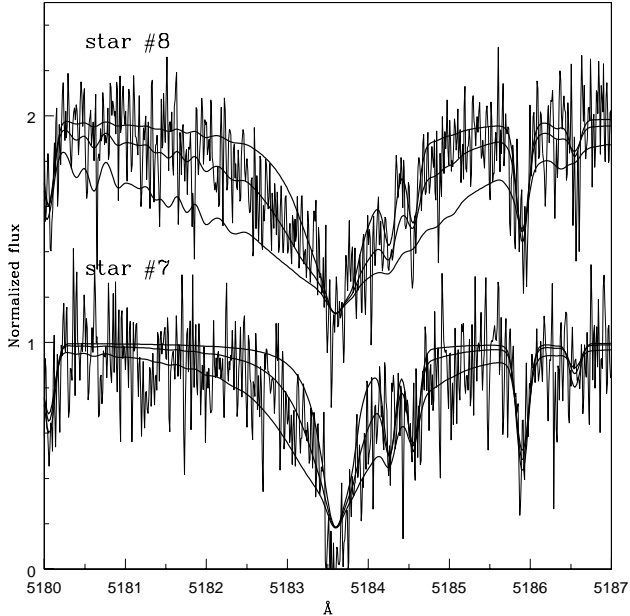}}
      \caption{The MgI line of stars \#7 and \#8, with overplotted (solid lines) synthetic profiles of 
               templates with corresponding temperatures. The plotted models are for $\log{g}$=3.0 
               (narrower, upper profiles), $\log{g}$=4.0 (middle lines), and $\log{g}$=5.0 (wider, lower 
               profiles). The spectra of star \#8 was vertically shifted adding a unit constant.}
       \label{fig_logg}
  \end{center}
\end{figure}

            A similar routine was prepared to estimate $\log{g}$, fitting the wings of three gravity-sensitive lines: 
            the CaI line at 6162 \AA~ (Edvardsson 1988, Katz et al. 2003), the redder line of the MgIb triplet at 5182 
            \AA (Kuijken et al. 1989), and the NaI doublet at 5890-5893 \AA~. The temperature was fixed to the 
            previously-determined value, interpolating the spectra using the synthetic library of Coelho et al. (2005), 
            while both solar metallicity and [Fe/H]=$-$0.5 were used. The three lines were fitted independently, 
            resulting in six gravity measurements per star. For stars \#4  and \#5 we obtained $\log{g}$=1.00$\pm$0.10 
            and 3.70$\pm$0.10 respectively, confirming again the reliability of the results. 
            In Figure~\ref{fig_logg} the MgI line of stars \#7 and \#8 is shown. As it can be seen, synthetic line 
            profiles change noticeably with gravity, and a rough estimate of stellar gravity can even be done by visual
            inspection.

            In order to fix the microturbulence velocity, we considered the equations available in the literature that 
            estimate this parameter as a function of T$_\mathrm{eff}$ and $\log{g}$. For star \#7 our choice fell on 
            the relation presented by Reddy et al. (2003), that was calibrated using stars of the same spectral type as
            the target and presents a small quoted error (0.15 km s$^{-1}$). Star \#8 is cooler than the temperature 
            range where this equation was calibrated. The relation proposed by Gratton et al. (1996) returns a much 
            lower $v_t$ for cool stars, but it was calibrated on a mix of dwarfs and giants in a large range of 
            temperatures, and the quoted uncertainty is much larger (0.30 km s$^{-1}$). As no approach was fully 
            satisfactory because both equations had a characteristic weakness, we decided to adopt the mean value as 
            the best estimate of $v_t$ for star \#8. However, different assumptions have only limited effects, because 
            both relations return a metallicity within 1$\sigma$ of the final estimate.

            The metallicity of stars \#7 and \#8 was measured from the EWs of the few iron lines for which a reliable 
            fit was possible. The number of available lines was small (about 15), but their fit was good and the 
            atmospheric parameters had already been determined. The rms of [Fe/H] from single lines was about 0.15 dex, 
            resulting in a formal error on the mean of about 0.04 dex. This was quadratically summed to the variations 
            of [Fe/H] induced by an increase/decrease of one parameter (T$_\mathrm{eff}$, $\log{g}$, $v_t$) by 1$\sigma$.

         \subsubsection{Reddening and distances}
            We finally compared the position of program stars in the T$_\mathrm{eff}$-$\log{g}$ plane with Yale-Yonsei 
            isochrones (Yi et al. 2003) of correct metallicity. The star position in this plane does not depend on 
            reddening or distance, but it can depend on age. For each star we adopted the isochrone that matched the 
            stellar parameters. However, this gives a reliable age estimate only for subgiant stars (\#5 and \#8), while
            at the tip of the RGB (star \#4) and in the MS (\#7) age is ill-defined because isochrones are too closely 
            spaced. For stars \#5 and \#8 we obtained an age estimate of 2.3$\pm$1.0 and 12$\pm$3 Gyr respectively, 
            where the errors were defined by the most extreme isochrones compatible with the 1$\sigma$ errors in 
            temperature and gravity. We calculated the color excess $E(V-I$) by comparison of observed 
            and theoretical colors. $E(V-I$) was then translated into $E(B-V$) using the relations found by Cardelli et al. 
            (1989). The two estimates of reddening usually differed because of observational errors, hence we adopted as 
            final value their weighted mean with associated error.  Results are presented in column~8 of 
            Table~\ref{spectradata}. Uncertainties on color excesses were calculated from propagation of errors on 
            observed and intrinsic colors, the latter being estimated as the variation induced by a change of 
            T$_\mathrm{eff}$ by $\pm 1\sigma$. In fact, we verified that the intrinsic color is minimally affected by the 
            errors on the other parameters ($\log{g}$, [Fe/H] and age). All but one of the values are very similar, and 
            the mean reddening is $E(B-V$)=0.28. The extinction map of Schlegel et al. (1998) returns $E(B-V$)=0.295 in 
            the direction of NGC\,6863, in excellent agreement with our estimate. This also confirms that the temperatures 
            adopted for program stars are substantially correct.

            From isochrones we also estimated the absolute magnitude of each object, and we calculated the distance 
            assuming $E(B-V$)=0.28 and A$_\mathrm{V}$=3.1\,$E(B-V$). The error on M$_\mathrm{V}$ was assumed as the 
            quadratic sum of the differences induced by the variation of each parameter (T$_\mathrm{eff}$, $\log{g}$,
            [Fe/H] and age) by 1$\sigma$, and the uncertainty on distance was calculated from propagation of errors.
            Results are presented in column~9 of Table~\ref{spectradata}. It is worth noticing that the distance of the 
            MS star \#7 is well-defined, while errors on surface gravity induce large uncertainties on the distance of 
            subgiants (\#5 and \#8).

            It is interesting to note that our distance and age for star \#5 agree well with the values cited by Dias 
            (2009) for NGC\,6863 (1200 pc, 3.47 Gyr), but this is true for this star only.
%
%
    \begin{figure}
     \centering
      \resizebox{\hsize}{!}{\includegraphics[]{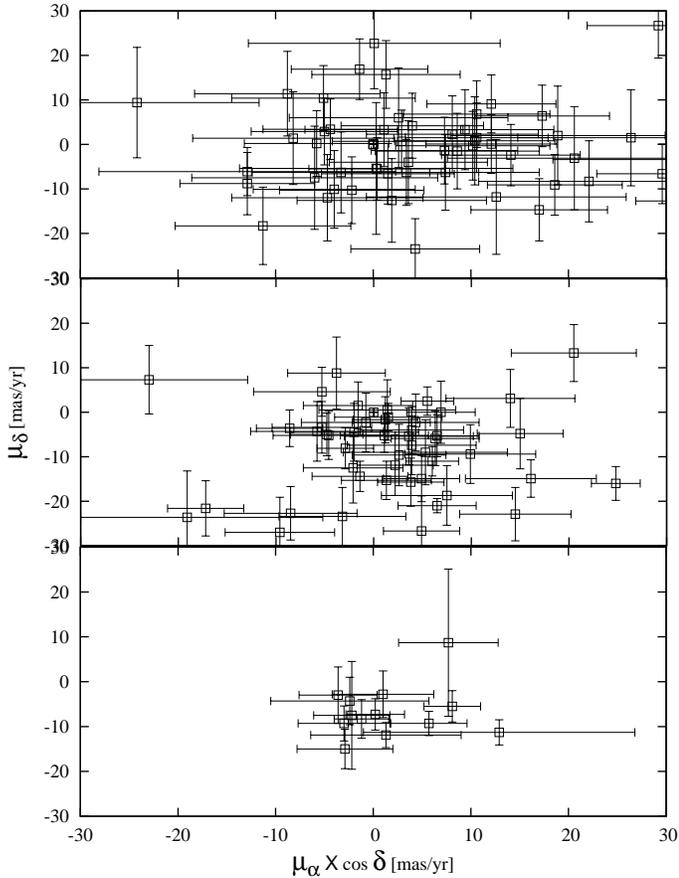}}
      \caption{Extended proper motion analysis. 
               {\it Top panel:} stars having 12 $ \leq K < 14$. 
               {\it Middle panel:} stars having 10 $ \leq K < 12$. 
               {\it Bottom panel:} stars brighter than $K = 10$.
              }
      \label{all}
    \end{figure}
%
%

   \subsubsection{Spectroscopic results: discussion}
      Metallicity, distance and radial velocity of star \#4 clearly show that it is a distant background star without 
      any connection with the others. At a height $z$=2.1 kpc from the Galactic plane and with intermediate metallicity,
      it appears as a typical thick disk red giant. To a lesser extent, a similar conclusion can be drawn for star \#8 
      also. Its properties are less extreme, but its RV is very different from those of the two remaining stars, and 
      its metallicity 
      is lower by more than 3$\sigma$. Stars \#5 and \#7, on the other hand, have very similar metallicity. Their RVs 
      differ by about 14 km s$^{-1}$, but this fact alone does not exclude that the two objects can be part of a stellar
      system, because they were observed in a single epoch only, and one or both could be binaries. It is worth pointing
      out that indeed a very high binary fraction is expected in OCRs (de la Fuente Marcos 1997, 1998), and among the 13
      confirmed members of NGC\,1901, to date the only spectroscopically-confirmed OCR, Carraro et al. (2007) measured at 
      any epoch an internal RV dispersion between 16 and 20 km s$^{-1}$. However, the spectroscopic distance of stars \#5 
      and \#7 still differ by 2.5$\sigma$. This is also consistent with the fact that there is no isochrone with 
      [Fe/H]=$-$0.08 
      and $E(B-V$)=0.28 that fits the stellar position of both stars in the CMD. This difference excludes the possibility 
      that they are part of the same stellar system. As a conclusion, the spectroscopic analysis reveals that the four 
      brightest stars of NGC\,6863 are not physically related; they are just a random alignment along the line of sight.
      In other words, they are an asterism.
%
%
    \begin{figure}
     \centering
      \resizebox{\hsize}{!}{\includegraphics[height=8cm]{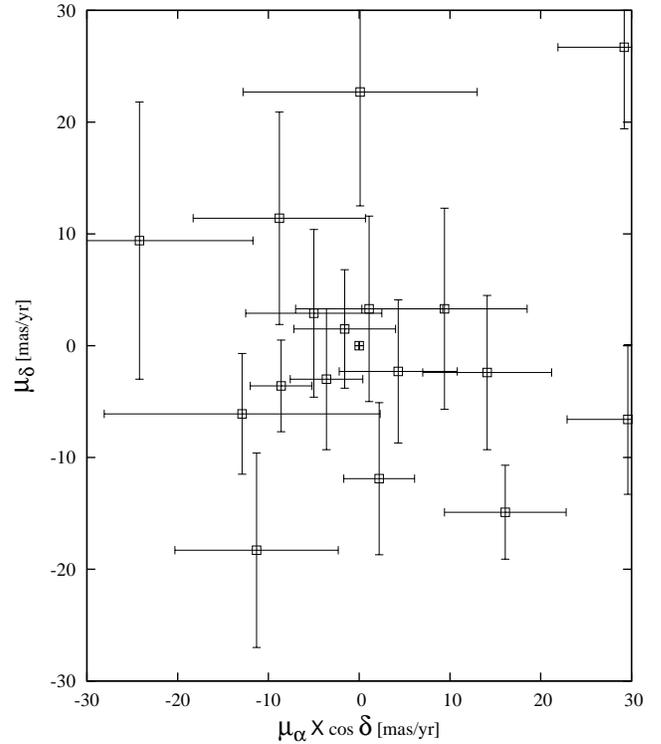}}
      \caption{Proper motion analysis. UCAC3 stars within 3 arcmin from the
               centre of the object.  
               }
      \label{centre}
    \end{figure}
%
%
   \section{Proper motions}
      Following Carraro et al. (2007), we extracted the proper motion 
      components in a 20-squared arcmin field around NGC~6863 from the UCAC3 
      catalogue (Zacharias et al. 2009), and constructed 3 vector-point 
      diagrams as a function of the $K$ magnitude (UCAC3 is cross-correlated 
      with 2MASS). The results are shown in Fig. \ref{all}, where the bottom 
      panel is restricted to stars having $K<$ 10, the middle panel to stars 
      in the range $10 \leq K < 12$, and the top panel to stars having 
      $12 \leq K < 14$. Stars in the lower panel appear to clump around 
      $\mu_{\alpha}$ = -2.0 [mas/yr], $\mu_{\delta}$ = -7.5 [mas/yr]. 
      Identifiable, consistent clumps are also observed in the other two 
      panels. Is this apparent clump the kinematic signature of NGC 6863? 

      Following Dias (2009) and restricting our analysis to the central
      region, 3 arcmin around the centre of the cluster, the results are
      rather different, see Fig. \ref{centre}. Clumping, if any, is observed 
      around $\mu_{\alpha}$ = -1.5 [mas/yr], $\mu_{\delta}$ = 1.5 [mas/yr].
      This diagram includes all the stars brighter than $K = 14$ with UCAC3 
      data within 3 arcmin of the centre of the object. In principle, we are 
      using data similar (they use UCAC2 not UCAC3) to those in Pavani \& Bica 
      (2007) but a direct comparison is not possible as
      they do not provide a diagram to illustrate the modulus of the proper 
      motion distribution for NGC 6863. They only indicate that its 
      characteristic median velocity is the highest in their open cluster 
      remnant sample, 137 km/s. In any case, preliminary visual inspection of
      Fig. \ref{centre} appears to confirm that the stellar population located 
      around the commonly accepted centre of NGC~6863 is kinematically 
      heterogeneous which is incompatible with open cluster status. In the 
      following section, this conclusion is explored by using a more detailed 
      statistical analysis.

      As for the four brightest stars discussed above, individual proper 
      motions are shown in Table \ref{pmdata}. Calculating the 
      Galactic space velocity for these stars as described in Johnson \&
      Soderblom (1987) with UCAC3 data in the Heliocentric reference frame 
      velocity $UVW$, we obtain results fully consistent with our previous 
      spectroscopic analysis. For star \#4, $U = 92$ km s$^{-1}$,  
      $V = -320$ km s$^{-1}$, and $W = -211$ km s$^{-1}$, which is  
      consistent with a halo star in a retrograde orbit. Star \#5 has 
      $U = 52$ km s$^{-1}$, $V = -14$ km s$^{-1}$, and $W = 28$ km s$^{-1}$.
      Similar numbers are found for star \#7 with $U = 51$ km s$^{-1}$,  
      $V = 2$ km s$^{-1}$, and $W = 11$ km s$^{-1}$. Star \#8 is unusual and
      it could be a thick disk member with $U = 25$ km s$^{-1}$, 
      $V = 58$ km s$^{-1}$, and $W = 206$ km s$^{-1}$.

\begin{table*}
\caption{Proper motion data for the brightest stars}
\label{pmdata}
\begin{tabular}{ccccc}
\hline
ID & $\alpha(J2000)$ & $\delta(J2000)$ & $\mu_{\alpha}$ UCAC3 & $\mu_{\delta}$ UCAC3 \\ 
 & ($^{h}$ $^m$ $^s$) & ($^{\circ}$ ' '') & [mas/yr] & [mas/yr] \\ 
\hline
4 & 20:05:08.50 & -03:33:22.7 &   2.2$\pm$3.9  & -11.9$\pm$6.8 \\
5 & 20:05:07.67 & -03:32:59.3 &  -8.6$\pm$3.4  &  -3.6$\pm$4.1 \\
7 & 20:05:06.07 & -03:33:29.3 & -12.9$\pm$15.2 &  -6.1$\pm$5.4 \\
8 & 20:05:06.78 & -03:33:26.6 & -49.9$\pm$6.1  &  35.5$\pm$6.0 \\
\hline
\end{tabular}
\end{table*}

   \section{An asterism in Aquila: statistical analysis}
      At this point, one may want to know how significant, statistically 
      speaking, is the kinematic evidence against NGC 6863 being a real open 
      cluster. It is true that the four brightest stars in the field of NGC 
      6863 do not appear to form a physical system but what if just one of 
      them is the brightest member of a real but poorly populated open 
      cluster like NGC 1901? We are facing a spatial data mining problem 
      where the goal is detection of a hypothetical over-density of kinematic
      nature (i.e. a small group of stars with similar proper motions) but in 
      addition, statistical testing must be performed in order to determine 
      whether the over-dense regions are significant. In order to distinguish 
      the clumps that are significant from those that are likely to have 
      occurred by chance we use Kulldorff's spatial scan statistic ($D_K$ 
      defined below, Kulldorff 1997). In the following, we assume data has 
      been aggregated to an $N \times N$ grid of squares, where each grid cell 
      is associated with a count $c_{ij}$ (number of stars with certain 
      properties, proper motions in this case) and an underlying population 
      $p_ {ij}$, with $i, j = 1,...,N$. Our goal is to find over-densities: 
      spatial regions where the counts are significantly higher than expected, 
      given the underlying population. Our objective is to search over all 
      square regions and find the region(s) with the highest density according 
      to a density measure. Kulldorff's statistic assumes that counts $c_{ij}$ 
      are generated by an inhomogeneous Poisson process with mean $q \ p_{ij}$, 
      where $q$ is the underlying expected value of the count to population
      ratio. We then calculate the logarithm of the likelihood ratio of two 
      possibilities: that $q$ is higher in the region than outside the region 
      or that it is identical inside and outside the region. For a cell with 
      count $c_{ij}$ and population $p_{ij}$, in a grid with total count $C$ 
      and population $P$, we can calculate:
      \begin{equation}
         D_{Kij} = c_{ij} \ \log \frac{c_{ij}}{p_{ij}} +
                   (C - c_{ij}) \ \log \frac{C - c_{ij}}{P - p_{ij}} -
                   C \ \log \frac{C}{P} 
                       \label{kulldorff}
      \end{equation}    
      if $c_{ij}/p_{ij} > C/P$, and $D_{Kij} = 0$ otherwise. Kulldorff (1997)
      proved that the spatial scan statistic is individually more powerful
      for finding a single significant region of unusually high density
      than any other test statistic. Unfortunately, performance of scan 
      statistics is affected by tessellation geometry (variation of cell size 
      and cell shape) as well as aggregation level (average cell size). A single
      actual cluster that is bisected by a large cell may appear either as 
      not significant or as two distinct clusters depending on whether the 
      large cell is included in the candidate zone (Kulldorff, Tango, \&
      Park 2003). In the following analysis, special care has been taken to 
      minimize tessellation effects by using multiple grid sizes and checking 
      for consistency. On the other hand, we have calculated $D_{Kij}$ for 
      each cell in the grid, the standard deviation, and plot 
      $(D_{Kij} -<D_{K}>)/\sigma$ using colour maps. To test the reliability 
      of our statistical approach we first apply it to a well established 
      OCR, NGC 1901, and then to our current object, NGC 6863.

      In order to carry out this analysis we use data from 2MASS.  The 2MASS 
      All Sky Catalog of Point Sources (Cutri et al. 2003) includes data for 
      both NGC 1901 and NGC 6863 and 20$\times$20 arcmin$^2$ areas around the 
      coordinates of the accepted centre of these objects were extracted. 
      Following Pavani \& Bica (2007) we use $E(B-V)$ = 3.03$\times E(J-H)$, 
      $A_J/A_V = 0.282$ and $A_J$ = 2.65$\times E(J-H)$ (Rieke \& Lebofsky 
      1985).

    \subsection{NGC 1901}
\begin{figure*}
     \centering
      \resizebox{\hsize}{!}{\includegraphics[height=7cm,angle=-90]{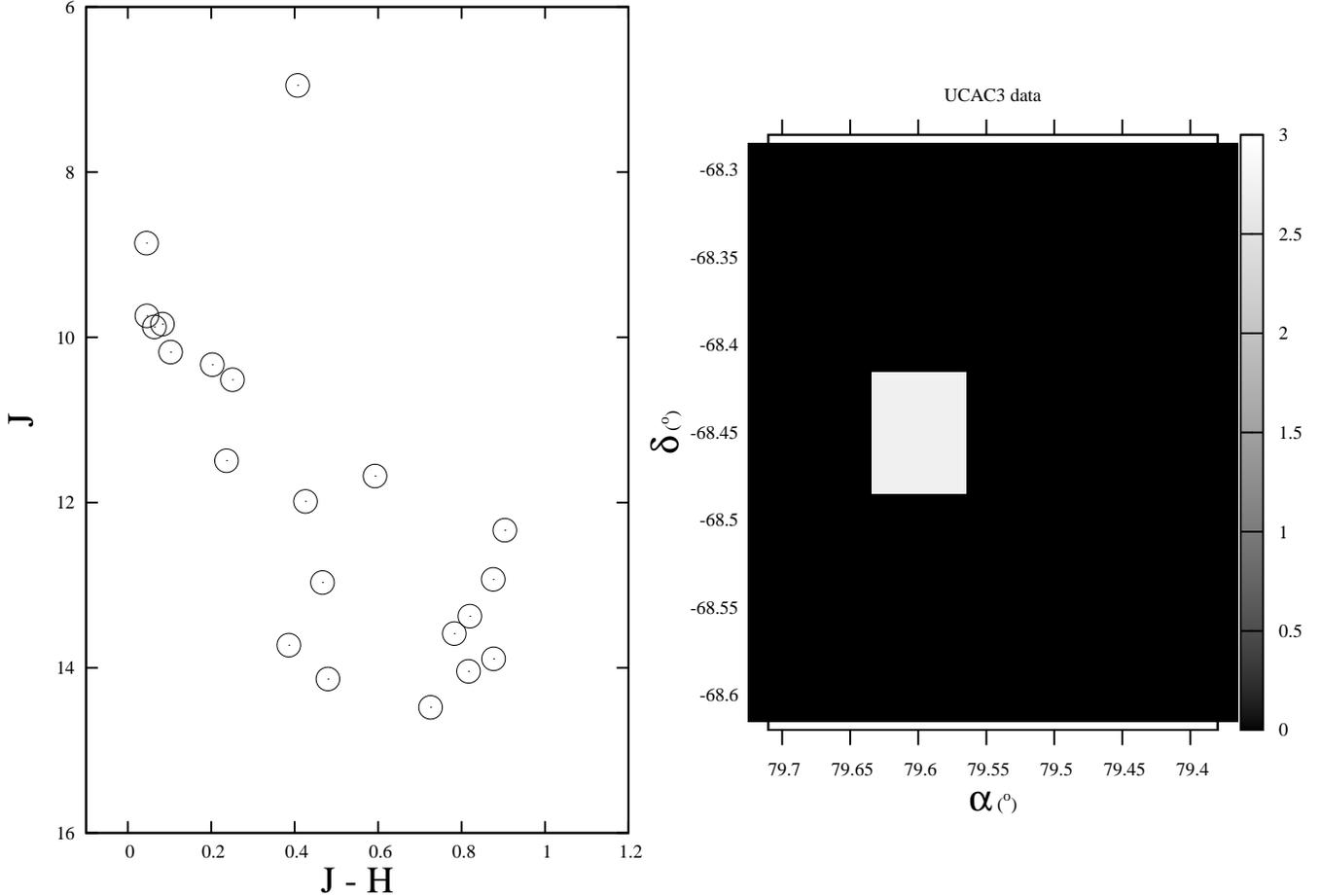}}
      \caption{NGC 1901. {\it Left panel:} 2MASS CMD for those stars with 
               proper motions in UCAC3 within 2$\sigma$ of ($\mu_{\alpha}, 
               \mu_{\delta}) = (1.3\pm1.5, 11.9\pm1.5)$ mas/yr. 
               {\it Right panel:} Kulldorff's spatial scan statistics for the 
               region around NGC 1901 (20$\times$20 arcmin$^2$ field, UCAC3 
               data). There is an obvious, statistically significant 
               ($\sim$3$\sigma$) over-density of stars with similar proper 
               motions centred at the accepted coordinates for this OCR.  
              }
      \label{kulNGC1901}
    \end{figure*}

         Using UCAC2 data, Carraro et al. (2007) concluded that the average 
         proper motions of this genuine OCR were ($\mu_{\alpha}, \mu_{\delta}) 
         = (1.7\pm1.3, 12.3\pm2.9)$ mas/yr. Repeating the analysis with UCAC3 
         data for the suspected NGC 1901 members compiled in Table 2 of that 
         paper we obtain ($\mu_{\alpha}, \mu_{\delta}) = (1.3\pm1.5, 
         11.9\pm1.5)$ mas/yr. This translates into $U = -22.4\pm8.4$ 
         km s$^{-1}$, $V = -3.0\pm1.7$ km s$^{-1}$, and $W = -1.9\pm2.8$ 
         km s$^{-1}$, which is consistent with other recent results (see 
         Bovy et al. 2009, Fig. 21). If we apply the spatial scan statistics 
         to the corresponding UCAC3 sample (20$\times$20-arcmin$^2$ around the 
         cluster centre) looking for over-densities within 2$\sigma$ of the 
         average proper motions, we obtain the plot in Fig. \ref{kulNGC1901} 
         (right panel). The effective size of the OCR is $\sim$3 arcmin 
         (Carraro et al. 2007); this is the grid cell size used in the 
         analysis. A statistically 
         significant over-density, $\sim$3$\sigma$, appears at the expected 
         position. The 2MASS CMD is also provided, $E(B-V)$ = 0.04 was used in 
         the reduction process. The CMD in Fig. \ref{kulNGC1901} is analogous 
         to the one in Pavani \& Bica (2007) even if a slightly different 
         $E(B-V)$ was used; see their Fig. 7, central panel. Photometry does 
         show a defined main sequence and the CMD hints of all the 
         characteristic properties of OCRs as predicted by $N$-body simulations
         (de la Fuente Marcos 1998): scarcity of low-mass members, high binary 
         fraction (a double main sequence is not unlikely), and overall sparse 
         population. These results together with the kinematically homogeneous
         signature confirm that the statistical analysis described above
         generates robust results. NGC 1901 is a true representative of the 
         terminal stage in the life of an open cluster, the OCR phase. Its 
         current properties are compatible with NGC 1901 being what remains of 
         a relatively small cluster with an original population of 500-750 
         stars (Villanova et al. 2004b, Carraro et al. 2007).     

%
%
%
%
  
%
%
    \begin{figure*}
     \centering
      \resizebox{\hsize}{!}{\includegraphics[height=7cm,angle=-90]{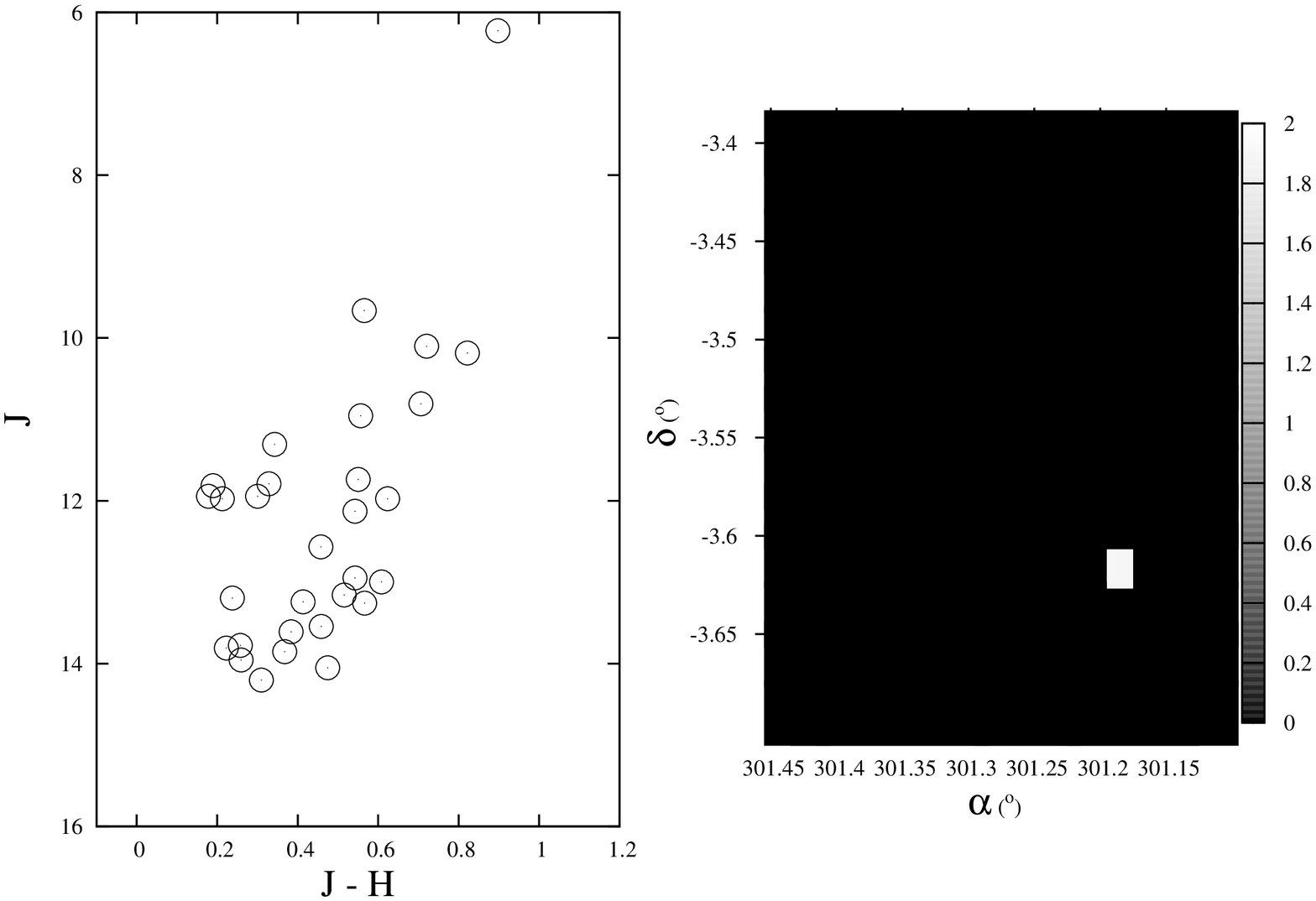}}
     \centering
      \resizebox{\hsize}{!}{\includegraphics[height=7cm,angle=-90]{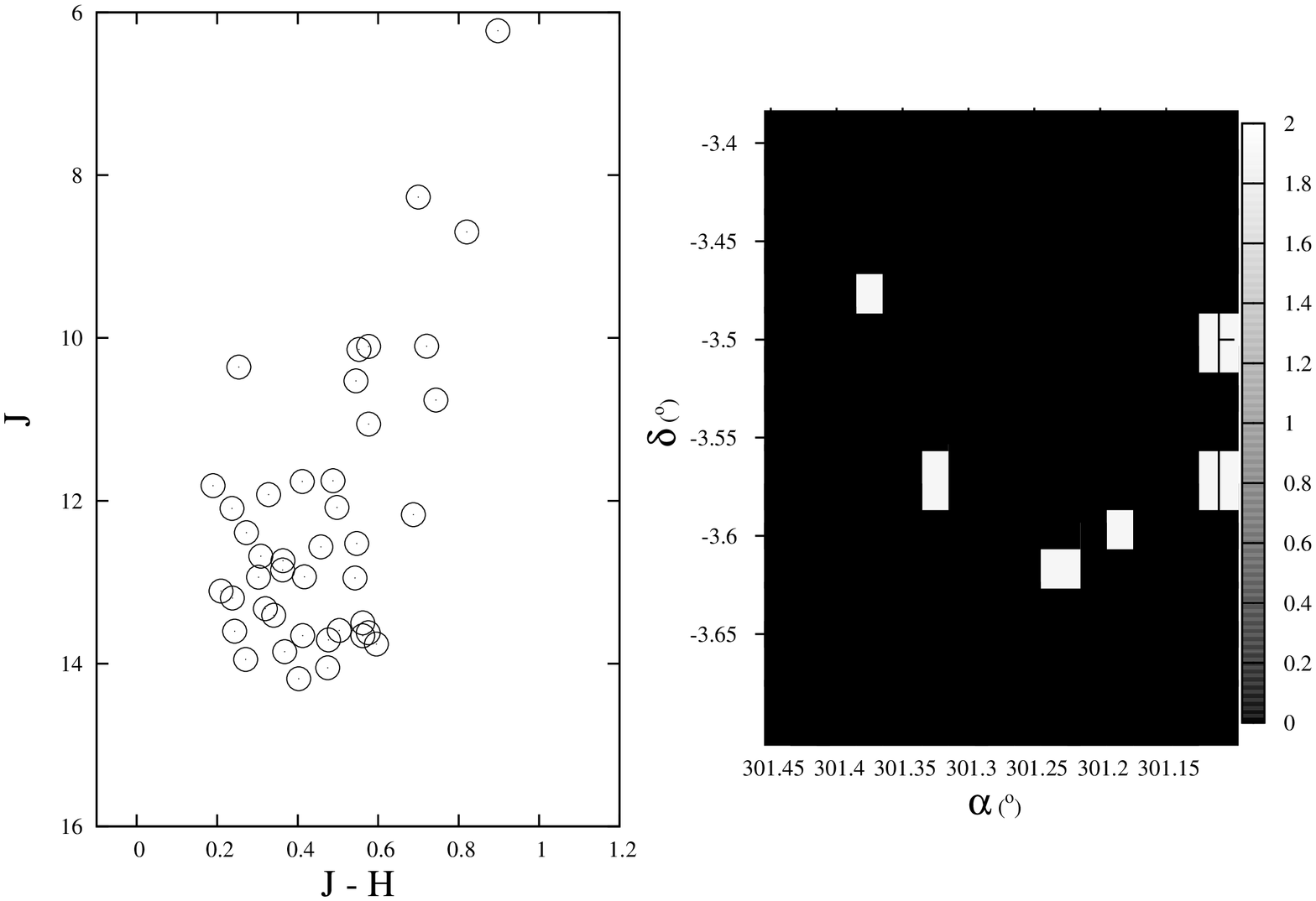}}
      \caption{Star \#4. {\it Top left panel:} 2MASS CMD for those stars with 
               proper motions in UCAC3 within 1$\sigma$ of ($\mu_{\alpha}, 
               \mu_{\delta}) = (2.2\pm3.9, -11.9\pm6.8)$ mas/yr. 
               {\it Top right panel:} Kulldorff's spatial scan statistics for 
               star \#4. No statistically significant clustering is visible 
               in this plot. 
               Star \#7. {\it Bottom left panel:} 2MASS CMD for those stars with 
               proper motions in UCAC2 within 1$\sigma$ of ($\mu_{\alpha}, 
               \mu_{\delta}) = (-12.9\pm15.2, -6.1\pm5.4)$ mas/yr. 
               {\it Bottom right panel:} Kulldorff's spatial scan statistics for 
               star \#7. There is no significant overdensity at the expected 
               location of NGC 6863. As for stars \#5 and \#8, no overdensity of 
               any kind is observed.
              }
      \label{kulstar47}
    \end{figure*}
%
%

      \subsection{The NGC 6863 asterism}
         If we repeat the same approach looking for over-densities associated 
         to the four brightest stars in the field of NGC 6863 we obtain Fig. 
         \ref{kulstar47}. Here, the effective size of the 
         hypothetical OCR is $\sim$1.5 arcmin (Pavani \& Bica 2007) and this 
         is the grid cell size used in the analysis. $E(B-V)$ = 0.28 was used 
         to plot the CMDs. No obvious over-density can be found at the 
         location of NGC 6863. It is, therefore, reasonable to conclude that 
         the actual presence of an open cluster, remnant or not, at the 
         location of NGC 6863 is very unlikely. The application of the 
         Kulldorf statistics shows that NGC 1901 is a real OCR, whereas it
         confirms that there is no evidence to suggest that NGC 6863 really 
         exists.

   \section{Conclusions}
      In this paper we have presented detailed photometric, spectroscopic and
      kinematic evidence against the classification of NGC 6863 as open cluster.
      Our results show that the four brightest stars commonly associated to 
      NGC~6863 form an asterism, a group of non-physically associated stars 
      projected together. The spectra of the four brightest stars in this field 
      clearly indicate that they are part of different populations. Their radial
      velocities are statistically very different and their spectroscopic 
      parallaxes are inconsistent with them being part of a single, bound 
      stellar system. Out of the four stars, only two of them have similar 
      metallicity. As for the underlying open cluster remnant described in 
      Pavani \& Bica (2007), our analysis of the existing data has demonstrated  
      convincingly that there is no base whatsoever to consider NGC 6863 as a
      possible OCR. The color magnitude diagrams for the field of 
      NGC~6863 do not even show any clear signature typical of actual open clusters.
      Spatial scan statistics using UCAC3 data strongly suggests that no 
      statistically significant, kinematically supported over-density is 
      present at the purported location of NGC~6863 although the proper motion 
      errors are rather large. This together with the other issues discussed 
      above force us to conclude that the presence of a star cluster of any 
      kind associated to the coordinates of NGC 6863 is highly unlikely.

   \begin{acknowledgements}
      CMB acknowledges the Chilean Centro de Excelencia en Astrof\'\i sica 
      y Tecnolog\'\i as Afines (CATA), and Francesco Mauro for very usefull
      discussions. GC deeply acknowledges the entire Asiago technical staff 
      for the kind night assistance over the whole duration of this project.
      In preparation of this paper, we made use of the NASA Astrophysics 
      Data System and the ASTRO-PH e-print server. This research has made
      use of the WEBDA database operated at the Institute of Astronomy of 
      the University of Vienna, Austria. This work also made extensive use of
      the SIMBAD database, operated at the CDS, Strasbourg, France. 
      This publication makes use of data products from the Two Micron All Sky
      Survey, which is a joint project of the University of Massachusetts and
      the Infrared Processing and Analysis Center/California Institute of 
      Technology, funded by the National Aeronautics and Space Administration
      and the National Science Foundation. 
   \end{acknowledgements}

\end{document}